\documentclass[aps,pra,twocolumn,superscriptaddress,floatfix,longbibliography]{revtex4}
\usepackage{multirow}
\usepackage{makecell}
\usepackage{amsfonts}
\usepackage{amssymb}
\usepackage{graphicx}
\usepackage{dcolumn}
\usepackage{bm}
\usepackage{amsmath}
\usepackage[colorlinks,linkcolor=magenta,citecolor=blue,urlcolor=blue]{hyperref}
\usepackage{changes}
\usepackage{braket} 
\usepackage{graphicx}
\usepackage{subfigure}
\usepackage{float}
\usepackage{color}

\begin{document}

\preprint{This line only printed with preprint option}
	
\title{Emergent strength-dependent scale-free mobility edge in a non-reciprocal long-range Aubry-Andr\'e-Harper model}

 \author{Gui-Juan Liu}
\affiliation{School of Physics, South China Normal University, Guangzhou 510006, China}

 \author{Jia-Ming Zhang}
\affiliation{School of Physics, South China Normal University, Guangzhou 510006, China}

\author{Shan-Zhong Li}
\email[Corresponding author: ]{shanzhongli88@gmail.com}
\affiliation{School of Physics, South China Normal University, Guangzhou 510006, China}

\author{Zhi Li}
\email[Corresponding author: ]{lizphys@m.scnu.edu.cn}

\affiliation {Key Laboratory of Atomic and Subatomic Structure and Quantum Control (Ministry of Education), Guangdong Basic Research Center of Excellence for Structure and Fundamental Interactions of Matter, School of Physics, South China Normal University, Guangzhou 510006, China} 
\affiliation {Guangdong Provincial Key Laboratory of Quantum Engineering and Quantum Materials, Guangdong-Hong Kong Joint Laboratory of Quantum Matter, Frontier Research Institute for Physics, South China Normal University, Guangzhou 510006, China}

\date{\today}
	
\begin{abstract}
We investigate the properties of mobility edge in an Aubry-Andr\'e-Harper model with non-reciprocal long-range hopping. The results reveal that there can be a new type of mobility edge featuring both strength-dependent and scale-free properties. By calculating the fractal dimension, we find that the positions of mobility edges are robust to the strength of non-reciprocal long-range hopping. Furthermore, through scale analysis of the observables such as fractal dimension, eigenenergy and eigenstate, etc., we show that four different specific mobility edges can be observed in the system. This paper extends the family tree of mobility edges and hopefully it will shed more light on the related theory and experiment.
\end{abstract}

\maketitle

\section{Introduction}
In 1958, Anderson first proposed the famous localization theory. The theory tells us that strong random disorder will destroy the statistical physical ergodicity of electrons, which makes the corresponding wave function decay exponentially with space~\cite{PWAnderson1958}. Anderson localization theory has drawn great attention in the past few decades because of not only the importance of the theory itself, but also the fact that many new devices have been developed based on it. Nowadays, the study of localization theory has gradually extended from solid materials to more different systems, such as photonic crystal~\cite{MStorzer2006,TSchwartz2007,JTopolancik2007}, waveguide array~\cite{YLahini2010,CThompson2010}, circuit system~\cite{SPai2019}, trapped ions~\cite{ABermudez2010}, ultracold atomic gases~\cite{JBilly2008,GRoati2008,AAspect2009,DHWhite2020}, etc. 

Then in 1970, scaling theory showed that an arbitrarily small disorder would lead to localization in a low-dimensional ($D<3$) system~\cite{EAbrahams1979,BHetenyi2021}. In the 3D case, the extended state and the localized state may coexist with an energy boundary, i.e., a mobility edge (ME), emerging between the two~\cite{NFMott1967,PALee1985,FEvers2008}. Past decades have seen great progress in the research on ME. Researchers have come to know that ME not only appears in random disordered systems, but may also appear in quasi-periodic disordered systems~\cite{SDasSarma1988,SDasSarma1990,JBiddle2010,YWang2020,JKoo1975,SAGredeskul1989,DNChristodoulides2003,TPertsch2004,GRoati2008,YLahini2009,GModugno2010,YEKraus2012,MVerbin2013,DTanese2014,MVerbi2015,SGaneshan2015,MLohse2016,SNakajima2016,PRoushan2017,XLi2017,FAAn2018,HPL2018,HYao2019,XDeng2019,SLonghi2019a,VGoblot2020,PWang2020,TLiu2020,AŠtrkalj2021,FAAn2021,LJZhai2021,NRoy2021,SRoy2021,SLonghi2021,WHan2022,TLiu2022,XXia2022,APadhan2022,SZLi2023,NSchreiber2015,HLi2023}. Unlike random disordered systems, quasi-periodic systems are of some limbo states between order and disorder. Because quasi-periodic models are easy to deal with theoretically (some even have analytical solutions~\cite{SDasSarma1988,SDasSarma1990,JBiddle2010,SLonghi2019a,YWang2020,SLonghi2021,TLiu2022,XXia2022,SGaneshan2015}), and easy to be realized experimentally~\cite{JKoo1975,SAGredeskul1989,DNChristodoulides2003,TPertsch2004,GRoati2008,YLahini2009,GModugno2010,YEKraus2012,MVerbin2013,DTanese2014,MVerbi2015,MLohse2016,SNakajima2016,PRoushan2017,FAAn2018,HPL2018,NSchreiber2015,HLi2023}. Nowadays they have gotten ever-increasing attention. One of the most typical quasi-periodic models is the 1D Aubry-Andr\'e-Harper (AAH) chain, which possesses the property of self-duality~\cite{SAubry1980}. Before (after) the critical point of phase transition, the system behaves as a pure extended phase (localized phase) without the emergence of ME~\cite{SAubry1980,SYJitomirskaya1999}. When the parameters are at the critical point, self-duality will work to ensure that real and reciprocal spaces have the same expression, and the corresponding eigenstates in the system exhibit the characteristics of multifractal states. 

Furthermore, energy-dependent traditional MEs have been successfully induced in quasi-periodic AAH models by introducing long-range (LR) hopping~\cite{JBiddle2010,XDeng2019,NRoy2021}, dimerized hopping~\cite{SRoy2021,WHan2022}, and modulating the quasi-periodic potential~\cite{SGaneshan2015,XLi2017,HPL2018,HYao2019,VGoblot2020,TLiu2020,YWang2020,AŠtrkalj2021,LJZhai2021,APadhan2022,SZLi2023}, etc. In addition to the traditional ME of deviding extended states and localized states, new types of non-traditional MEs have also been explored in recent years. So far, it has been found that the introduction of $p$-wave paraing~\cite{SZLiPwave,JWang2016,MYahyavi2019,TLiu2021a,JFraxanet2022}, quasi-periodic hopping~\cite{FLiu2015,JCCCCestari2016,LWang2017,LZTang2021,TXiao2021,TLiu2021b,XCZhou2023}, LR hopping~\cite{ZXu2021,NRoy2021,DPeng2023,MG2023} into the AAH model can induce the non-traditional ME involving multifractal states. 

On the other hand, because of its excellent performance in describing dissipation or non-equilibrium process, wide attention has been paid on the study of non-Hermitian properties~\cite{ZGong2018,SYao2018,VMMartinezAlvarez2018,FKKunst2018,FSong2019a,CHLee2019a,KYokomizo2019,CHLee2019,SLonghi2019,HJiang2019,DWZhang2020a,DWZhang2020b,ZYang2020,YLiu2020,NOkuma2020,KZhang2020,KKawabata2020,CScheibner2020,YYi2020,LLi2020,THelbig2020,AGhatak2020,SWeidemann2020,LXiao2020,LJZhai2020a,LJZhai2020b,DSBorgnia2020,LLi2021,LJLang2021,LXiao2021,EJBergholtz2021,XZhang2022,KZhang2022,QLin2022a,QLiang2022,QLin2022b,ZXGuo2022,HZLi2023,BLi2023,CXGuo2023,SZLi2024,XJYu2024}. Many new phenomena previously not found in Hermitian systems have been unveiled one after another~\cite{EJBergholtz2021,XZhang2022}. Non-Hermitian skin effect of non-reciprocal (NR) model, a typical phenomenon that only exists in non-Hermitian systems, has been extensively studied in recent years~\cite{SYao2018,VMMartinezAlvarez2018,FKKunst2018,KYokomizo2019,CHLee2019,SLonghi2019,HJiang2019,NOkuma2020,KZhang2020,KKawabata2020,CScheibner2020,YYi2020,LLi2020,THelbig2020,AGhatak2020,SWeidemann2020,LXiao2020,DSBorgnia2020,LLi2021,LXiao2021,KZhang2022,QLin2022a,QLiang2022,QLin2022b,BLi2023,CXGuo2023}. It has been noticed that the corresponding eigenstates in systems with non-Hermitian skin effect are very sensitive to the boundary conditions~\cite{SYao2018}. 

So far, most of the research focuses on the AAH model with LR hopping or the non-Hermitian LR ordered model. More precisely, recent study reported that, AAH models with LR hopping can produce a new type of strength-dependent ME~\cite{XDeng2019}, while another study on non-Hermitian systems pointed out that NRLR hopping leads to the emergence of scale-free localized states in the system~\cite{YCWang2023}. Scale-free localized states behave as skin states in small system size, and extended states in large system size. Based on the above studies, it is natural to wonder whether there can be a ME featuring both strength-dependent and scale-free properties in an AAH model with NRLR hopping. 

To answer this question, this paper will devote to the study of NRLR AAH model.
\begin{figure}[thbp]
\centering	
\includegraphics[width=8cm]{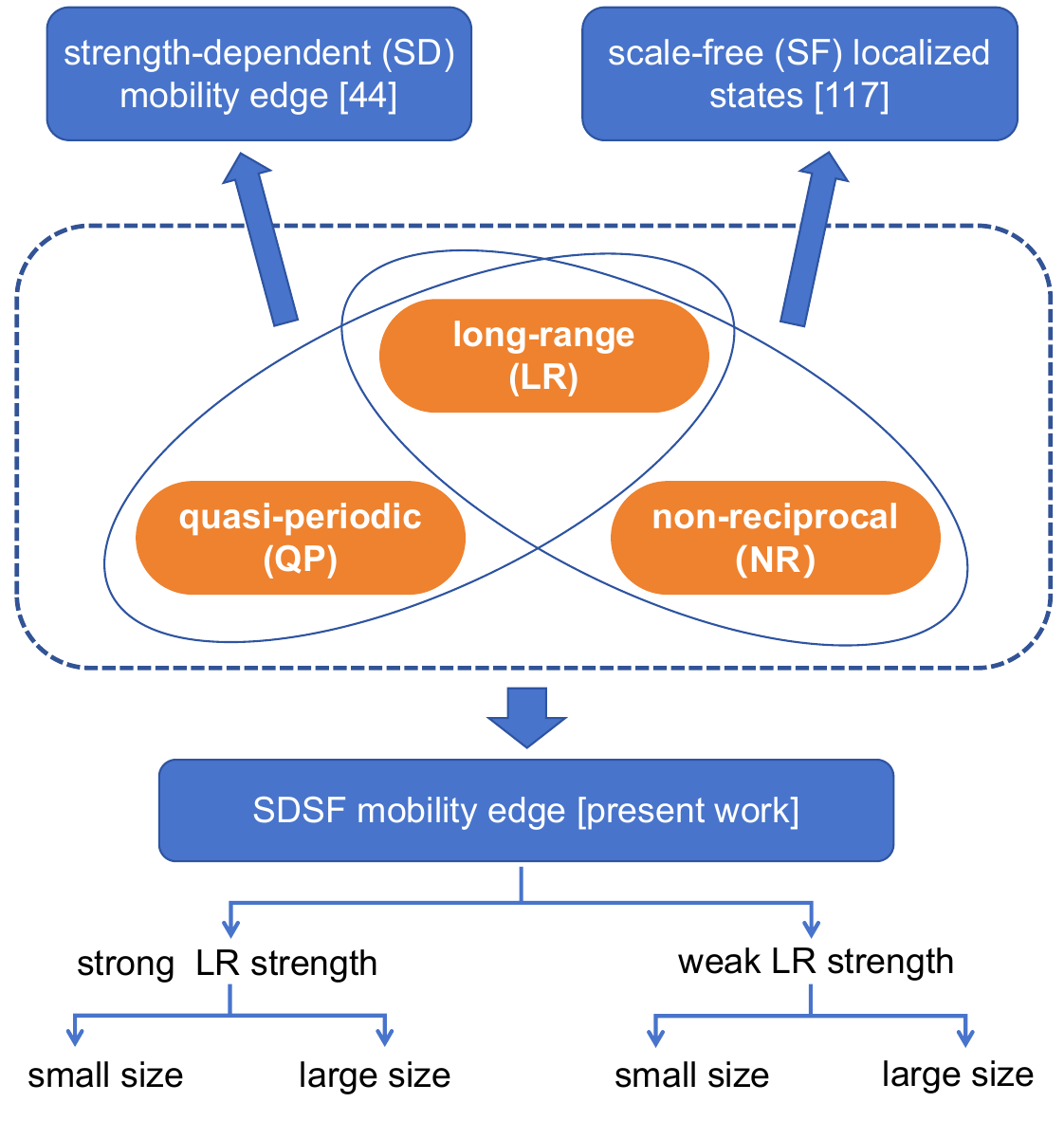}
\caption{The scheme for implementing SDSF MEs.}\label{F00}
\end{figure}

The rest of the paper is structured as follows. In Sec.~\ref{II}, we give a brief introduction to the theoretical model. In Sec.~\ref{III} and Sec.~\ref{IV}, we discuss properties of the system’s ME with strong and weak LR hopping, respectively. The results of this paper and that in Hermitian case are compared and analyzed in Sec.~\ref{V}. The experiment realization has been discussed in Sec.~\ref{VI}. The main conclusions are summarized in Sec.~\ref{VII}.

\section{MODEL}\label{II}
We start at a NRLR Aubry-Andr\'e-Harper model. The corresponding Hamiltonian reads
\begin{equation}\label{Hami}
\begin{aligned}
&H=H_{\text{NRLR}}+H_{\text{QP}}, \\
&H_{\text{NRLR}}=\sum_{i<j}^{}\frac{-J_{L}}{|i-j|^{a}}c_{i}^{\dagger}c_{j}+\sum_{i>j}^{}\frac{-J_{R}}{|i-j|^{a}}c_{j}^{\dagger}c_{i}, \\&H_{\text{QP}}=\sum_{j=1}^{L}\lambda\cos(2\pi\alpha j+\theta)c_{j}^{\dagger}c_{j},
\end{aligned}
\end{equation}
where $H_{\text{NRLR}}$ and $H_{\text{QP}}$ correspond to the NRLR hopping and the quasi-periodic potential, respectively. $c_{j}$ ($c_{j}^{\dagger}$) denotes the fermionic annihilation (creation) operator at site $j$, $J_{L}$ ($J_{R}$) is the leftward (rightward) hopping strength, $L$ is the system size, and $a$ is the LR parameter, which determines the strength of LR hopping. The on-site potential is dominated by the quasi-periodic strength $\lambda$ and the quasi-periodic parameter $\alpha$, as well as the random phase $\theta$. In the computing process, we set $\alpha=\lim_{v\rightarrow\infty}F_{v-1}/F_{v}=(\sqrt{5}-1)/2$, which can be obtained from the Fibonacci numbers $F_{v+1}=F_{v}+F_{v-1}$, where $F_{0}=0$ and $F_{1}=1$. 

\begin{figure*}[thbp]
\centering	
\includegraphics[width=18cm]{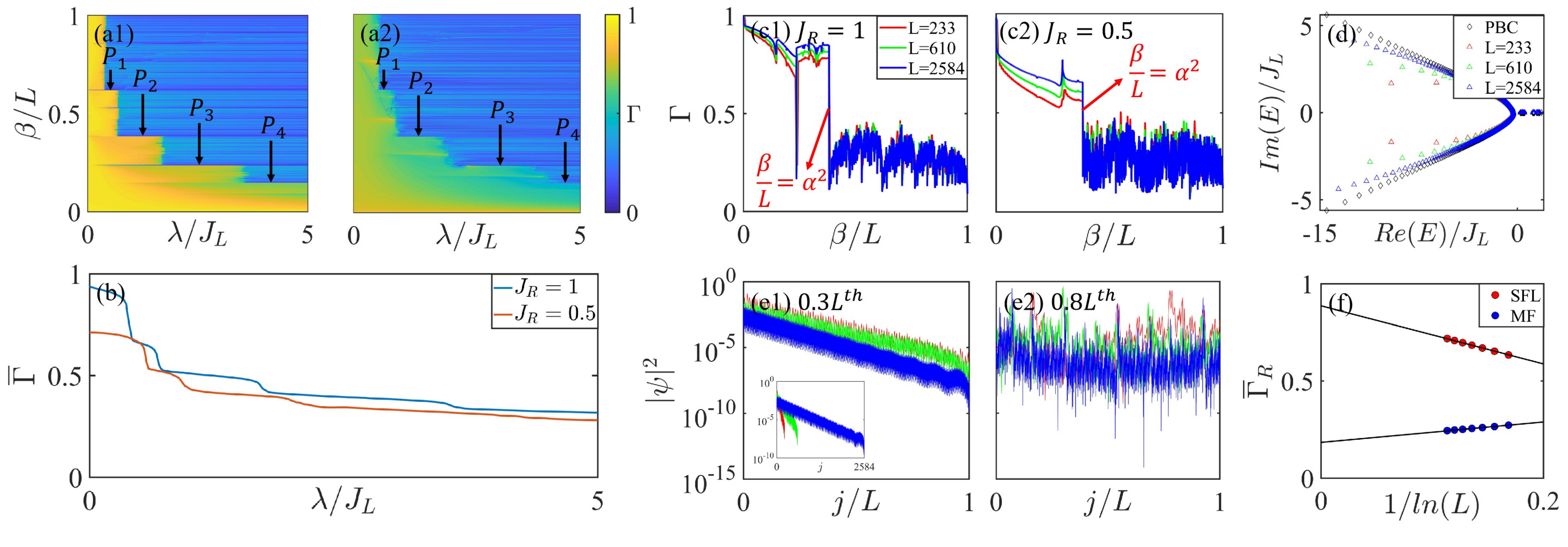}
\caption{The fractal dimension $\Gamma$ of all eigenstates versus $\lambda$ with $J_R=1$ (a1) and $J_R=0.5$ (a2). (b) The corresponding average fractal dimension $\overline{\Gamma}$. The scale effect of $\Gamma$ with respect to the level index $\beta/L$ at $\lambda=1.26$ for $J_R=1$ (c1) and $J_R=0.5$ (c2). (d) The eigenenergy in the complex plane for different sizes with $\lambda=1.26$, where the black diamond coresponds to PBCs with $L=2584$. (e1) and (e2) show the density distribution of $0.3L^{th}$ and $0.8L^{th}$ eigenstates in log scale. (f) is  the average fractal dimension $\overline{\Gamma}$ versus $L$ for SFL and multifractal (MF) regions. System size $L=610$ in (a) and (b). Throughout, we set $a=0.5$.}\label{F11}
\end{figure*}

\begin{figure*}[htbp]
\centering	
\includegraphics[width=18cm]{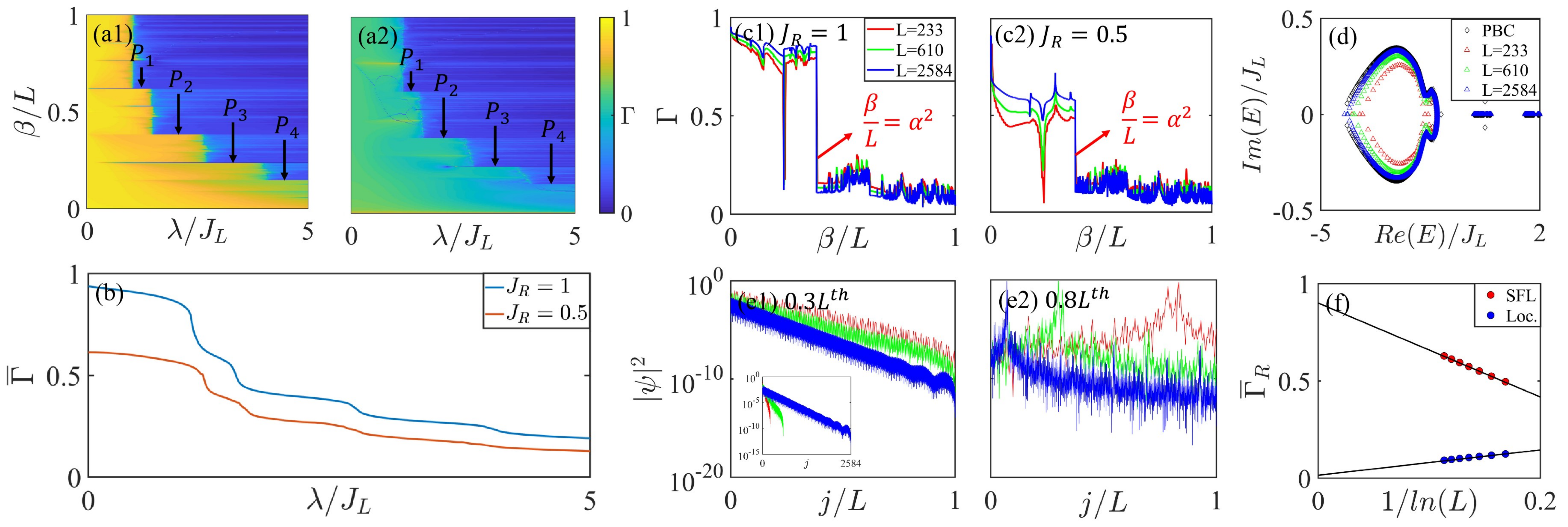}
\caption{The fractal dimension $\Gamma$ of all eigenstates versus $\lambda$ with $J_R=1$ (a1) and $J_R=0.5$ (a2). (b) The corresponding average fractal dimension $\overline{\Gamma}$. The scale effect of $\Gamma$ with respect to the level index $\beta/L$ at $\lambda=2.08$ for $J_R=1$ (c1) and $J_R=0.5$ (c2). (d) The eigenenergy in the complex plane for different sizes with $\lambda=2.08$, where the black diamond coresponds to PBCs with $L=2584$. (e1) and (e2) show the density distribution of $0.3L^{th}$ and $0.8L^{th}$ eigenstates in log scale. (f) is  the average fractal dimension $\overline{\Gamma}$ versus $L$ for SFL and localized (Loc.) regions. System size $L=610$ in (a) and (b). Throughout, we set $a=1.5$.}\label{F22}
\end{figure*}

Without loss of generality, we fix $\theta=0$ and set $J_{L}=1$ as the unit of energy. When $J_{R}=1$, the Hamiltonian reduces to the reciprocal Hermitian case~\cite{XDeng2019}. In this case, both strong ($a<1$) and weak ($a>1$) LR hopping will cause a step-like phase transition from $P_1$ phase to $P_4$ phase with the increase of $\lambda$[see in Fig.~\ref{F11}(a1) and Fig.~\ref{F22}(a1)]. The corresponding position of ME will appear at $P_s=\alpha^{s}L$, where $s=1,~2,~3,~4$~\cite{XDeng2019}. Under the condition of $a<1$, the ME serves as a boundary to separate the extended state from the multifractal state, while under the condition of $a>1$, it will separate the extended state from the localized state.

\section {The SDSF MOBILITY EDGE FOR the case of STRONG NRLR HOPPING: $a < 1$}\label{III}
The fractal dimension, as the key observable to judge the localization properties and phase transition of quasi-perodic systems, is defined as~\cite{FWegner1980,HAoki1983,HAoki1986,MJanssen1998,ADMirlin2006,FEvers2008,ADeLuca2014,XDeng2016}
\begin{equation} \label{D}
\Gamma_{\beta}=-\lim_{L\to\infty}\frac{\ln\xi_{\beta}}{\ln L},
\end{equation}
where $\xi_{\beta}=\frac{\sum_{j=1}^{L}|\psi_{\beta,j}|^4}{[\sum_{j=1}^{L}|\psi_{\beta,j}|^2]^2}$ is the inverse participation ratio and $\psi_{\beta,j}$ is the amplitude at the $j^{th}$ site for the $\beta^{th}$ eigenstate. The $\Gamma$ approaches 1 and 0 for the extended and localized states, respectively, while $0 <\Gamma< 1$ for the multifractal states. 

First, let's discuss the strong NRLR case ($a<1$). By fixing the parameter $a=0.5$, we compute the fractal dimensions $\Gamma$ of all eigenstates as a function of $\lambda$ in the Hermitian ($J_R=1$) and the non-Hermitian ($J_R=0.5$) conditions, respectively~[see Fig.~\ref{F11}(a)]. One can see clearly that MEs of both Hermitian and non-Hermitian cases exhibit a step-like transition from $P_{1}$ to $P_{4}$, resulting in a step-like change in the localization properties. Note that, the $\Gamma$ of the extended state decreases with the introduction of non-Hermitian, further discussion will show that this implies the emergence of a new type state, i.e., scale-free localized (SFL) state. 
To further illustrate the trend of the step-like phase transition, we calculate the average fractal dimension $\overline{\Gamma}=\frac{1}{L}\sum_{\beta=1}^{L}\Gamma_{\beta}$ and plot them in Fig.~\ref{F11}(b). Blue and red line correspond to the Hermitian and non-Hermitian cases, respectively. The results show that for different $\lambda$, the fractal dimension $\overline{\Gamma}$ always decreases, and the gap between different phase regions ($P_{1}$-$P_{4}$) also narrows, which leads to the gradual flattening of the step-like phase transition.

Furthermore, we analyze the key quantities such as fractal dimensions, eigenvalues and eigenstates under different system sizes by scaling theory. 
We choose $\lambda = 1.26$, which corresponds to the center of the $P_2$ region, to discuss the ME. In Fig.~\ref{F11}(c), we plot the distribution of fractal dimension in different system sizes.
One can observe that for both Hermitian and non-Hermitian cases, the $\Gamma$ decreases abruptly at the energy index $\beta/L=\alpha^2\approx 0.382$, suggesting that the NR hopping does not change the position of the ME. On the one hand, for the Hermitian limit ($J_{R}=1$), one can find that in the region $\beta/L<\alpha^2$, $\Gamma$ tends to $1$ with an increasing system size, suggesting that the region is extended~\cite{FEvers2008}. Moreover, in the region $\beta/L>\alpha^2$, the $\Gamma$ is independent of system size, which indicates that the region is multifractal. This means, MEs of the Hermitian case are the boundary of extended states and multifractal states. On the other hand, for the NRLR case ($J_{R}=0.5$), while the corresponding fractal dimension $\Gamma$ in the region of $\beta/L<\alpha^2$ also increases with an increasing system size, the value magnitude of $\Gamma$ becomes smaller than the Hermitian case. Through further analysis one can find that in the region where $\beta/L>\alpha^2$, the fractal dimension $\Gamma$ remains independent of system size, which means the corresponding eigenstates are the multifractal states.

In order to better show the system's localization properties and the difference between open boundary condition (OBC) and periodic boundary condition (PBC), we show the eigenvalues in the complex plane for different system sizes [see Fig.~\ref{F11}(d)]. One can find that the eigenvalues in the region of $\beta/L<\alpha^2$ depend on the system size, and gradually converging to the case of the PBC. This is a solid evidence of SFL states, where the eigenvalues of OBC gradually converge to that of PBCs (or $L\rightarrow\infty$)~\cite{XDeng2019}. In the region of $\beta/L>\alpha^2$, the corresponding eigenvalues are always real, which are independent of the system size and boundary conditions. The scale-free property of the $\beta/L<\alpha^2$ region is also reflected in the eigenstates [see Fig.~\ref{F11} (e)]. Fig. 1(e1) shows the eigenstate corresponding to $\beta=0.3L$. It can be seen that the corresponding wave function is exponentially localized at the left boundary in the case of small size. Then, as the system size increases, the localization length will be protected, so that the distribution of the wave function will gradually be close to the extended state. Since SFL states exhibit an exponential profile $|\psi_{j}| \propto e^{-j\xi/L}$, where $j$ is the site index, $\xi$ is the localization length, $\xi/L$ is the decay strength, and is proportional to the system size $L$ for arbitrary $L$, so that the rescaled probability profiles $L|\psi(j/L)|$ show the scale-invariant behavior~\cite{YCWang2023}. Therefore, SFL states behave as skin states in small system size, and extended states in large system size. In other words, $\xi/L$ will be smaller as the system size becoming larger. Although as shown in Fig.~\ref{F11} (e1), it seems that the larger the size, the faster the decay. But that's actually not the case, for we normalize the system size. When the system size is not normalized, i.e. the horizontal axis directly uses the site index as shown in the inset in Fig.~\ref{F11} (e1), the distribution of the $0.3L^{th}$ eigenstate has a size-dependent localization length, and the eigenstate will become more and more extended as the system size increases, indicating the eigenstates in the region $\beta/L<\alpha^2$ are SFL states with size-dependent localization lengths. By a similar analysis, one can find that the eigenstates (the $0.8L^{th}$) remain multifractal in the multifractal region. 

Since the system has well-defined MEs, one can define the average fractal dimension 
\begin{equation}
\overline{\Gamma}_{R}=\sum_{R}\frac{1}{L_{R}}\Gamma_{R}
\end{equation}
for different regions to fit the thermodynamic limit, where $R=\mathrm{SFL}$ or $\mathrm{MF}$, which correspond to the eigenstates within the regions of $\beta/L<\alpha^{2}$ and $\beta/L>\alpha^{2}$, respectively. $L_{R}$ is the total number of eigenstates in region $R$. As shown in Fig.~\ref{F11}(f), we find that the average fractal dimension $\overline{\Gamma}_{R}$ for the SFL region increases with an increasing $L$, eventually converging to about $0.9$. While $\overline{\Gamma}_{R}$ does not strictly reach $1$, it is sufficient to illustrate the extended properties of the region. In contrast, the $\overline{\Gamma}_{R}$ in the multifractal region converges to a finite value. Therefore, the NRLR case for $a<1$ exhibits the coexistence of extended and multifractal states under the condition of thermodynamic limit. The MEs between them, which is the same to the Hermitian case, locate at $\beta_{c}/L=\alpha^2$.

\renewcommand{\arraystretch}{1.7} 
\begin{table*}[ht]
\caption{Comparison of mobility edges between reciprocal and non-reciprocal long-range AAH models}

\begin{tabular}{|c|cc|cccc|}
\hline
                   & \multicolumn{2}{c|}{Reciprocal long-range AAH~\cite{XDeng2019}}    & \multicolumn{4}{c|}{NRLR case(present work)}                                                                                                                  \\ \hline
Long-rang parameter & \multicolumn{1}{c|}{~~~~~~~$a<1$~~~~~~~}  & \multicolumn{1}{c|}{$a>1$} & \multicolumn{2}{c|}{$a<1$}  & \multicolumn{2}{c|}{~~$a>1$~~} \\
\hline
System size        & \multicolumn{2}{c|}{Independent of $L$}                 & \multicolumn{1}{c|}{$L \ll \infty$} & \multicolumn{1}{c|}{$L \rightarrow \infty$} & \multicolumn{1}{c|}{$L \ll \infty$} & \multicolumn{1}{c|}{~~$L \rightarrow \infty$~~} \\ \hline
Mobility edge      & \multicolumn{1}{c|}{\makecell{\\Extended \\+\\ Multifractal\\~~}}           & \multicolumn{1}{c|}{\makecell{\\Extended \\+\\ Localized\\~~}}              & \multicolumn{1}{c|}{\makecell{\\Skin Localized \\+\\ Multifractal\\~~}}                            & \multicolumn{1}{c|}{\makecell{\\Extended \\+\\ Multifractal\\~~}}                      & \multicolumn{1}{c|}{\makecell{\\Skin Localized \\+\\ Localized\\~~}}                            & \multicolumn{1}{c|}{\makecell{\\Extended \\+\\ Localized\\~~}}                      \\ \hline
\end{tabular}

\label{Tab1}
\end{table*}

\section{The SCALE-FREE MOBILITY EDGE FOR the case of Weak NRLR: $a>1$}\label{IV}
Let us turn to the weak NRLR case ($a > 1$). Under such circumstances, the ME as the boundary separates the extended and localized states in the Hermitian case ($J_{R} = 1$) and also undergoes a transition from the $P_1$ to the $P_4$ phase with an increasing $\lambda$~\cite{XDeng2019}. 

A similar analysis to the previous section is as follows.
In Fig.~\ref{F22}(a) we show the fractal dimensions $\Gamma$ versus $\lambda$ for $J_{R}=1$ and $J_{R}=0.5$, and both cases exhibit step-like localization phase transitions. Similar to the case of $a=0.5$, non-Hermitian version does not change the position of the MEs. This implies that in the case $a>1$ there will be another scale-free ME that separates SFL and Anderson localized states. In Fig.~\ref{F22}(b), we show the average fractal dimension $\overline{\Gamma}$ as a function of $\lambda$ for different $J_R$. Compared to the Hermitian case, $\overline{\Gamma}$ generally decreases, and the phase transition between different $P_{s}$ becomes smoother.

Then, we fix $\lambda=2.08$ (center of $P_2$ region) to discuss the emergent scale-free ME. In Fig.~\ref{F22}(c), we show the $\Gamma$ of all eigenstates at different system sizes. In both Hermitian and non-Hermitian cases, the ME is at $\beta/L=\alpha^2$ and is not affected by NR hopping. For $J_{R}=1$ (see Fig.~\ref{F22}(c1)), one can see that in the region where $\beta/L<\alpha^2$, as the system size increases, $\Gamma$ approaches 1, corresponding to extended states. In the region where $\beta/L>\alpha^2$, as the system size increases, $\Gamma$ approaches 0, corresponding to localized states. This indicates that the MEs separate the extended and localized states. As shown in Fig.~\ref{F22}(c2), the $\Gamma$ increases with system size increasing in the region where $\beta/L<\alpha^2$. Similar to the case of $a=0.5$, the $\Gamma$ remains small for finite sizes. Further scaling analysis reveals that the corresponding states are SFL states. In the region where $\beta/L>\alpha^2$, the $\Gamma$ approaches 0 as the system size increases, indicating that it is a localized region. 

The scale-free properties of the $\beta/L<\alpha^2$ region can also be characterized by the eigenvalue and the distribution of eigenstate. As shown in Fig.~\ref{F22}(d), one can see the gradual convergence of the eigenvalues in the $\beta/L<\alpha^2$ region to the PBC as the system size increases.  The distribution of the $0.3L^{th}$ eigenstate has a size-dependent localization length [see Fig.~\ref{F22}(e1)], and the eigenstates will become more and more extended as the system size increases. For the region $\beta/L>\alpha^2$, as shown in Fig.~\ref{F22}(d)(e2), both the eigenvalues and the eigenstates exhibit size-independent properties.

Finally, in Fig.~\ref{F22}(f) we conduct the scaling analysis of the average fractal dimension $\overline{\Gamma}_{R}$ of the two regions. The results reveal that by interpolating to the thermodynamic limit, the $\overline{\Gamma}_{R}=0$ for the localized region, while the $\overline{\Gamma}_{R}$ approaches $1$ for the SFL region, indicating the emergent scale-free ME that separates SFL states from Anderson localized states.

\section {discussion}\label{V}
The main finding of this paper is a new type of ME, which depends on both system size and LR strength. The difference of MEs between the reciprocal LR and NRLR case has been summarized in Table~\ref{Tab1}.

For the Hermitian case, when the LR effect is strong (weak) $a<1$ ($a>1$)[see Appendix ~\ref{A}], the MEs lie at the energy index $P_{s}=\alpha^{s}L$ separating the extended and multifractal (localized) states. Fortunately, the position formula obtained for the Hermitian case still holds for the non-Hermitian case. For the non-Hermitian case, the ME remains at the position of the energy indicator $P_{s}$. However, unlike the Hermitian version, the extended region will be replaced by the SFL state with a size-dependent localized length. To be more specific, when the system size is small, the SFL will localize on the boundary as a skin localized state. As the size increases the localization length increases (with a constant proportion of the system size), and eventually at system size infinity, it will become an extended state. Therefore, for $a<1$ ($a>1$), the ME will separate the SFL state and the multifractal (localized) state.

\section {EXPERIMENT REALIZATION OF THE SDSF MOBILITY EDGE}\label{VI}
Quantum simulators that can simulate AAH models, such as ultracold atoms~\cite{NSchreiber2015,GRoati2008,HPL2018}, optical waveguide arrays~\cite{SAGredeskul1989,DNChristodoulides2003,TPertsch2004}, and superconducting circuits~\cite{PRoushan2017,HLi2023}, can be considered as potential experimental platforms to realize the phenomena explored in this paper. We propose the experimental scheme for observing SDSF mobility edge by taking the ultracold atomic gas as an example. So far, based on $^{87}$Rb and $^{39}$K atomic gases, the standard AAH model has been successfully realized in bichromatic optical lattices experiments~\cite{NSchreiber2015,GRoati2008,HPL2018}. In concrete terms, the first step is to cool down and obtain an ensemble of Bose-Einstein condensates in a harmonic trap. In the second step, the interaction among atoms is adjusted to zero through Feshbach resonance technique. Finally, the cold atoms are placed into a one-dimensional bichromatic optical lattice tube. By controlling the depth and energy ratios of the bichromatic lattice, the quasi-periodic potential and the hopping strength can be precisely adjusted. 

In addition to the above experiment, it is also necessary to induce a long-range power-law hopping to realize the model in this paper. Experimentally, there are several ways to achieve long-range hopping in ultracold atomic lattice system. For example, in a recent report, the dynamics of a one-dimensional Bose gas with power-law decaying hopping amplitudes was studied after an abrupt reduction of the hopping range~\cite{WSDias2017,PRicherme2014,PJurcevic2014}. After obtaining the AAH model with long-range hopping, the non-reciprocal hopping can be readily obtained by manipulating the auxiliary laser and finally, the quantum simulation of the model discussed in this paper can be realized~\cite{ZGong2018,FSong2019a}. As for detection, it is necessary to detect the density distribution of the wave function under certain parameters by absorption imaging techniques. The phase of the system can be determined by the distribution of localized, extended and skin states, etc. indicated by the density distribution.


\section {CONCLUSIONS}\label{VII}
In conclusion, the ME properties of 1D NRLR AAH model are explored in this paper. The rest of the phase diagram spanned by the exponent $a$ and the hopping strength $J_{R}$ is discussed in Appendix ~\ref{B}. The results show that a new type of scale-free ME can be induced by introducing the full-space NRLR hopping into the AAH model. In addition, we find that the NR effect has little effect on the position of the ME, but has a great effect on the phase segmented by the ME, i.e., the extended region corresponding to the Hermitian case can be transformed to the size-dependent SFL region. Since the SFL state has a size-dependent localization length, the wave function in the system exhibits properties of skin localized state at small sizes but properties of extended state at large sizes. This results in four different MEs in the system, namely, Skin Localized + Multifractal, Extended + Multifractal, Skin Localized + Localized and Extended + Localized. Finally, several typical observables (such as the fractal dimension, eigenvalue and eigen wave function of the system) are analyzed by scale theory, so as to show from different perspectives that the size-dependent ME does appear in the system.

\begin{acknowledgments}
This work was supported by the Guangdong Basic and Applied Basic Research Foundation (Grants No.2021A1515012350) and the National Key Research and Development Program of China (Grant No.2022YFA1405300).
\end{acknowledgments}


\appendix
\section{The critical properties of $a=1$}\label{A}

\begin{figure}[hp]
	\centering 
	\includegraphics[width=8.5cm]{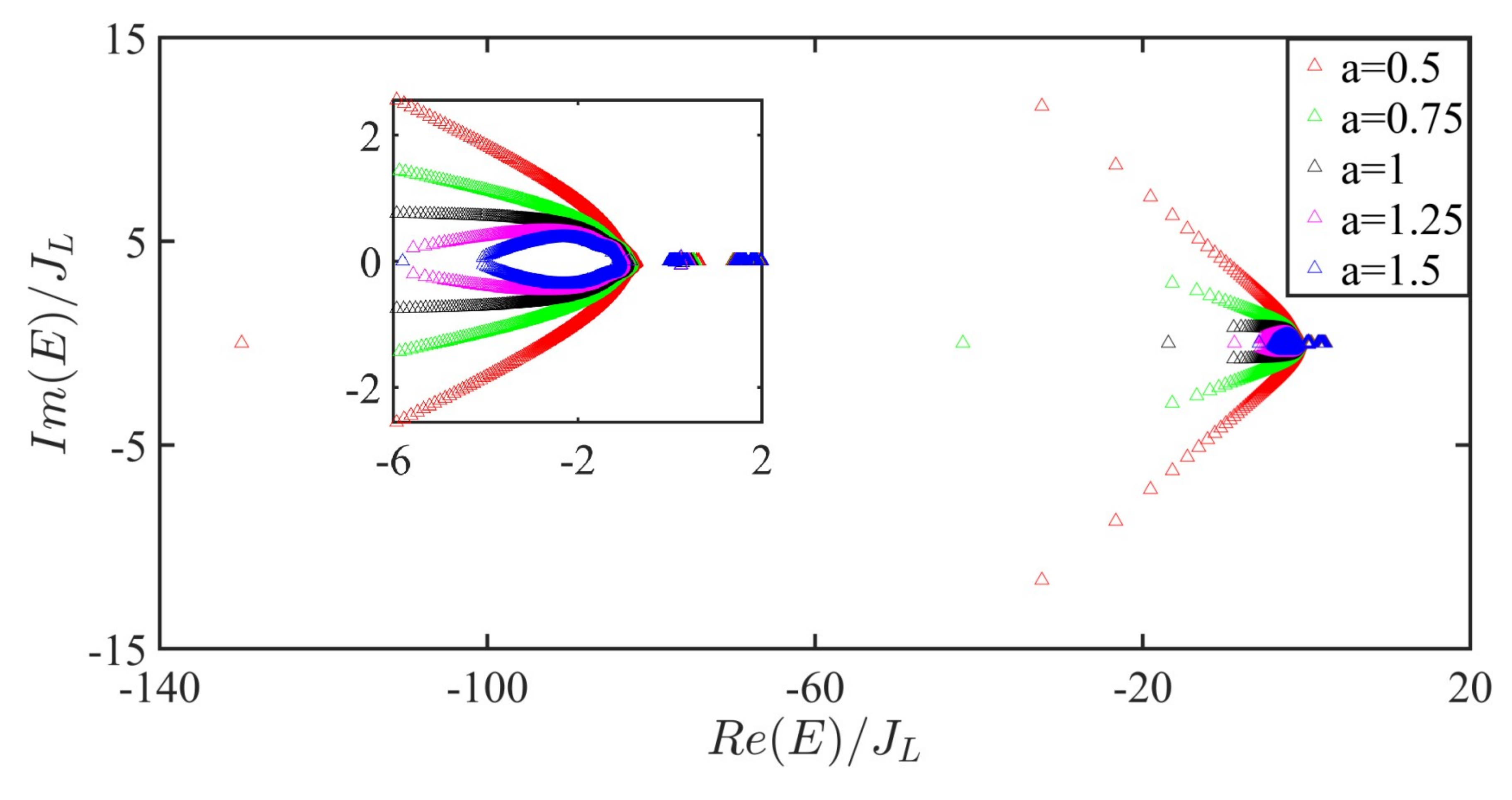} 
	\caption{Energy spectrum for different LR parameter $a$ under PBCs with $L=2584$. Throughout, we set $\lambda=1.8$.}
	\label{F44}
\end{figure}

The conclusions come from Ref.\cite{XDeng2019}, which provides a comprehensive exposition of the phase diagram of the long-range quasi-periodic model by numerical calculations, showing that $a<1(a>1)$ has mobility edges separating the extended state and the multifractal critical (localized) state. To show these conclusions more clearly, we have added corresponding description of the critical characteristics of the system in the case of $a=1$. To be more specific, we provide the energy spectrum to exhibit the case $a=1$ is truly the critical value(see Fig.~\ref{F44}).

As shown in Fig.~\ref{F44}, in the case of $a>1$, the energy spectrum in the complex plane will have a closed ring-like structure. That is to say, there will be two points of intersection with the real axis. In this case, the left point of intersection will move in the negative direction as parameter $a$ decreases. The position of the left intersection point will tend to  $-\infty$ for the case of $a=1$, which means only one intersection point left with the real axis in the energy spectrum and therefore cannot produce the above-mentioned closed ring-like structure. Note that, under the condition of $a>1$, the imaginary part of the spectrum remains always a finite value. Then in the case of $a \leq 1$, the spectrum will take on a half-open structure, and the opening will become larger with the decrease of $a$. In this case, the imaginary part of the energy will gradually diverge rather than stay at a finite value, as shown in Fig.~\ref{F44}.

\section{More details of different parameters $a$ and $J_{R}$}\label{B}

{We discuss the cases of different parameters $a$ and the hopping strength $J_{R}$. As shown in Fig.~\ref{F55}, we exhibit Hermitian $J_{R}=J_{L}$ in Fig.~\ref{F55}(a1)-(d1) and non-Hermitian $J_{R} \neq J_{L}$ in Fig.~\ref{F55} (a2)-(d2). 

\begin{figure}[hp]
	\centering 
	\includegraphics[width=8cm]{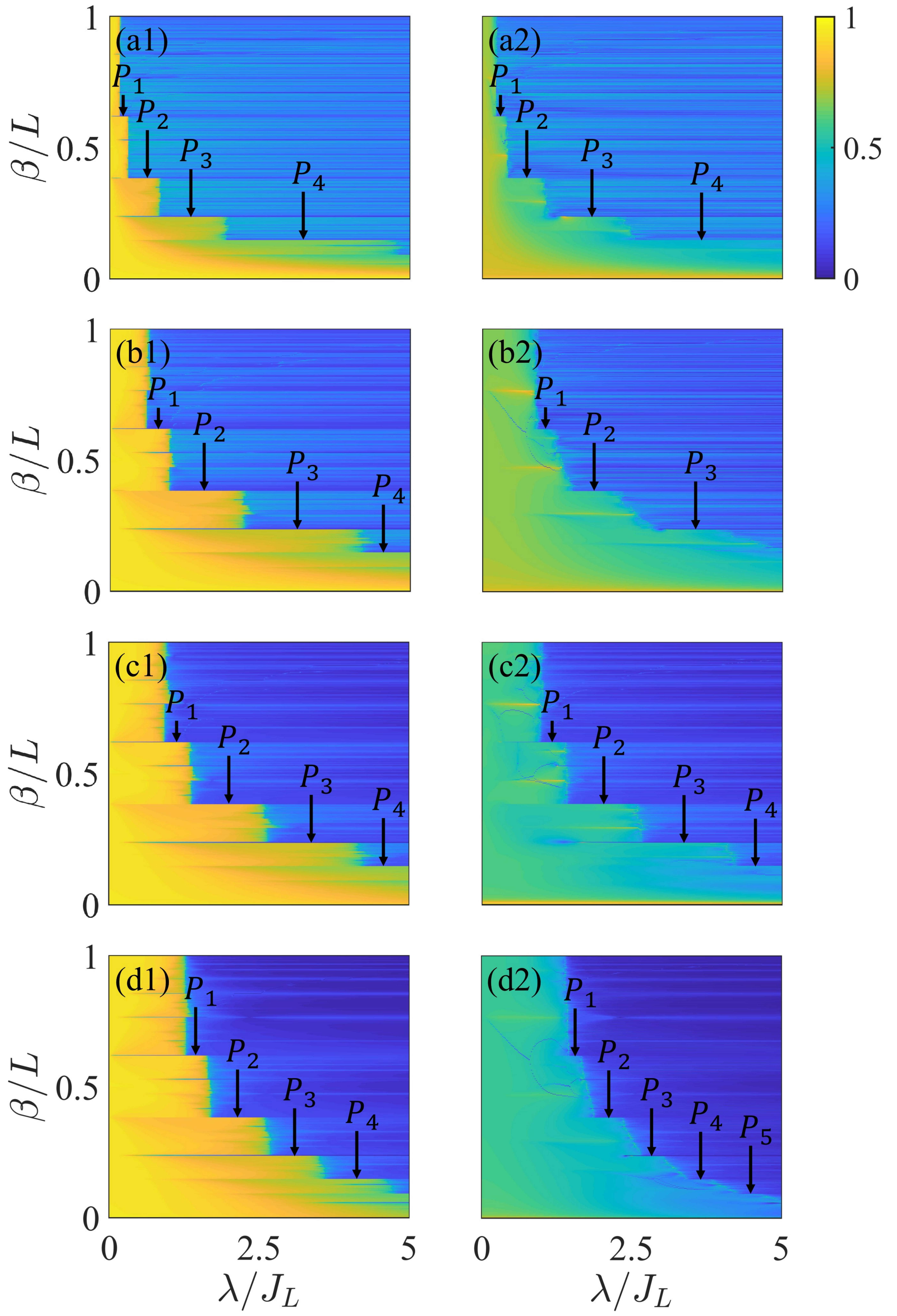} 
	\caption{(a) The fractal dimension $\Gamma$ of all eigenstates versus $\lambda$ when $a=0.2, J_{R}=1$ (a1), $a=0.2, J_{R}=0.8$ (a2), $a=0.8, J_{R}=1$ (b1), $a=0.8, J_{R}=0.2$ (b2), $a=1.3, J_{R}=1$ (c1), $a=1.3, J_{R}=0.8$ (c2), $a=2, J_{R}=1$ (d1), $a=2, J_{R}=0.2$ (d2). Throughout, we set $L=610$.}
	\label{F55}
\end{figure}

The results are similar to those discussed in the main text, in that the introduction of non-Hermitian does not change the position of the mobility edge $P_{s}$, but makes the step transition between different $P_{s}$ phases smooth. In addition, the fractal dimension of the eigenstates in the extended region decreases and transforms into SFL states under OBCs. When the parameter $a < 1 (a > 1)$, the ME separates the skin states and multifractal (localized) states in small sizes, and separates the extended states and multifractal (localized) states in large sizes.

\end{document}